\documentclass{pasj01}
\usepackage{natbib}

\Received{$\langle$reception date$\rangle$}
\Accepted{$\langle$acception date$\rangle$}
\Published{$\langle$publication date$\rangle$}
\newcommand{\J}{J0453.6$-$6829}

\begin{document}

\title{ High resolution X-ray study of supernova remnant J0453.6$-$6829 with unusually high forbidden-to-resonance ratio}
\author{Yosuke \textsc{Koshiba},$^{1}$ Hiroyuki \textsc{Uchida},$^{1}$ Takaaki \textsc{Tanaka},$^{2}$ Yuki \textsc{Amano},$^{1}$ Hidetoshi \textsc{Sano},$^{3}$ and Takeshi Go \textsc{Tsuru}$^{1}$}%
\altaffiltext{$^{1}$}{Department of Physics, Kyoto University, Kitashirakawa Oiwake-cho, Sakyo, Kyoto, Kyoto 606-8502, Japan }
\altaffiltext{$^{2}$}{Department of Physics, Konan University, 8-9-1 Okamoto, Higashinada, Kobe, Hyogo 658-8501, Japan }
\altaffiltext{$^{3}$}{National Astronomical Observatory of Japan, Mitaka, Tokyo 181-8588, Japan }
\email{koshiba.yosuke.b25@kyoto-u.jp}


\maketitle

\begin{abstract}
Recent high-resolution X-ray spectroscopy has revealed that several supernova remnants (SNRs) in the Large Magellanic Cloud (LMC) show unusually high forbidden-to-resonance ($f/r$) line ratios.
While their origin is still uncertain and debated, most of these SNRs have asymmetric morphology and/or show evidence of interaction with dense material, which may hint at the true nature of the anomalous $f/r$ ratios.
Here we report on a detailed spectral analysis of an LMC SNR \J \ with the Reflection Grating Spectrometer (RGS) onboard XMM-Newton.
We find that the $f/r$ ratio of O$\emissiontype{VII}$  ($=1.06^{+0.09}_{-0.10}$) is significantly higher than expected from the previously-reported thermal model.
The spectrum is fairly explained by taking into account a charge exchange (CX) emission in addition to the thermal component. 
Analyzing archival ATCA \& Parkes radio data, we also reveal that H\emissiontype{I} cloud is possibly interacting with \J.
These results support the presence of CX in \J, as the origin of the obtained high $f/r$ ratio.
Although a contribution of the resonance scattering (RS) cannot be ruled out at this time, we conclude that CX seems more likely than RS considering the relatively symmetric morphology of this remnant.

\end{abstract}

%\linenumbers

\section{Introduction}\label{sec:intro}
Plasma diagnostics of astrophysical objects using line ratios of He-like ions \citep[i.e., $G$ and $R$ ratios; ][]{Ga69} will become a part of mainstream in the upcoming era of high-resolution X-ray spectroscopy.
Recent grating observations revealed that several supernova remnants (SNRs) have anomalous line ratios of O$\emissiontype{VII}$, in which the forbidden line intensity relative to the resonance line  (hereafter, $f/r$) is significantly higher than expected for an ordinary thermal plasma \citep[e.g.,][]{Ka11, Uc19}.
While their physical origin is still under debate, two interpretations have mainly been argued: charge exchange (CX) and resonance scattering (RS), both of which were predicted to occur in SNRs by previous calculations \citep[e.g.,][]{La04, Ka95}.
The presence of the CX emission and/or RS effect would hinder accurate plasma diagnostics, and, more importantly, these physical processes themselves work as useful probes to obtain key information such as collision and turbulent velocities.
It is therefore required to know physical conditions and surrounding environments in which CX and/or RS can occur, that is, to reveal  the origin of the anomalously high $f/r$ ratios of O$\emissiontype{VII}$ found in SNRs.

On the basis of observations with the Reflection Grating Spectrometer (RGS) onboard XMM-Newton, previous  analyses of SNRs in the Large Magellanic Cloud (LMC) indicate that several of them tend to have relatively high $f/r$ ratios of O$\emissiontype{VII}$; N23 \citep{Br11}, N49 \citep{Am20}, and N132D \citep{Su20}.
They are located in a dense ambient medium as suggested by radio-line \citep[e.g.,][]{Ba97, Sa17} or infrared observations \citep[e.g.,][]{Wi06}, which may hint at the cause of the anomalous spectral features. 
In this context, we found that no detailed spectroscopy has been performed so far for the middle-aged LMC SNR \J, although this remnant is in a dense environment similar to N23 and N132D according to \citet{Wi06}.

\J \ is a relatively compact ($\sim2\arcmin$ in diameter) remnant of a core-collapse explosion \citep{Lo09, Mc12}, containing a  pulsar wind nebula (PWN) at the center of the shell \citep{Ga03}.
In addition to synchrotron radiation from the PWN, \citet{Mc12} indicated that the X-ray spectrum of \J \ obtained with Chandra is well explained by a shock-heated interstellar medium (ISM).
A similar conclusion was reached by \citet{Ha12}, who performed multi-frequency observations of \J, including X-ray band with XMM-Newton.
In their spectral fit, the forbidden line of O$\emissiontype{VII}$ is seemingly higher than a normally expected thermal model.
In this paper, we thus revisit the RGS data with particular attention to the He-like lines, in conjunction with an H$\emissiontype{I}$ observation around the remnant, in order to investigate the relation between the $f/r$ ratio and  surrounding environment of \J.
Errors are given at the 68\% confidence level throughout the paper.
We assume the distance to \J \  to be 50~kpc \citep{Pi13}.

\section{Observation and data reduction}\label{sec:obs}

\begin{figure}[h]
	\begin{center}
	\includegraphics[width=0.4\textwidth]{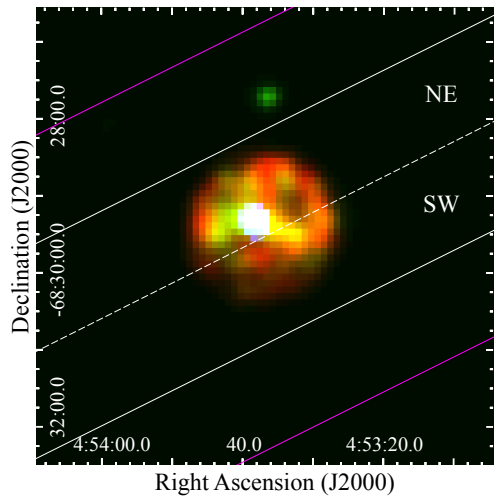}
	\end{center}
	\caption{
	True-color image of \J \ obtained with EPIC (MOS and pn).
	Red, green, and blue correspond to the energy bands of 0.3--1.2~keV (soft), 1.2--2.0~keV (medium), and 2.0--8.0~keV (hard), respectively.
	The magenta lines indicate the cross dispersion width of the RGS ($5\arcmin$).
	The spectra and background are extracted from the region sandwiched by the white lines and the region above the source region sandwiched by the magenta and white lines, respectively.
	The dashed line shows how the source region is divided for a spatially-resolved analysis.
	}
	\label{fig:full_img}
\end{figure}

\J \ was observed  with XMM-Newton on 2001 March 29 (Obs.~IDs~0062340101 and 0062340501).
Since one of the data sets (Obs.~ID~0062340501) was  affected by large soft-proton flares, we present results only from Obs.~ID~0062340101.
The raw data were processed with the XMM Science Analysis Software (SAS) version 18.0.0 and the calibration data files released in 2020 June.
In the following spectral analysis, we combine RGS1 and RGS2 data with MOS spectra. 
After discarding periods of background flares, we obtained MOS and RGS data with effective exposure times of $\sim6$~ks and $\sim20$~ks, respectively.
We do not analyze second order spectra because of their poor statistics.

\section{Analysis and results}\label{sec:analysis}
Figure~\ref{fig:full_img} shows a background-subtracted true-color image of \J \ taken by EPIC (MOS and pn).
We extracted RGS spectra by limiting the cross-dispersion width so as to cover the whole of \J.
MOS spectra were obtained from the entire region of the remnant.
Off-source regions in the field of view (FOV) were used to extract background spectra for each instrument.
We simultaneously fitted the unbinned RGS and MOS spectra using SPEX version 3.06.01 \citep{Ka96}, applying a maximum likelihood method, W-stat \citep{Wa79}.
Throughout this analysis, the hydrogen column density ($N_{\rm{H}}$) of the Galactic absorption was fixed to $6 \times 10^{20}$~cm$^{-2}$ \citep{Di90} and that of the LMC was left free.
We referred to \citet{Ru92} for the elemental abundances of the LMC.

\begin{figure*}[h]
	\begin{center}
	\includegraphics[width=0.8\textwidth]{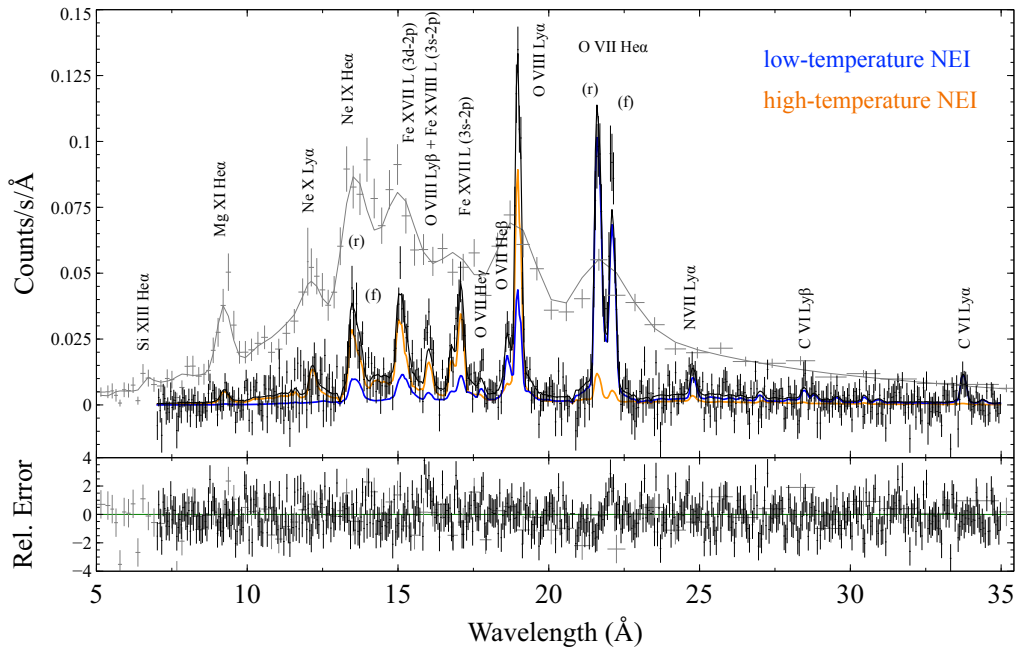}
	\end{center}
	\caption{
		RGS1$+$2 (black) and MOS1 (gray) spectra of \J.
		The colored solid curves indicate the contributions of the low-temperature (blue) and high-temperature NEI (orange) components.
		The bottom panel shows residuals from the best-fit model.
	}
	\label{fig:neij2_spectrum}
\end{figure*}

\begin{figure}[h]
	\begin{center}
	\includegraphics[width=0.35\textwidth]{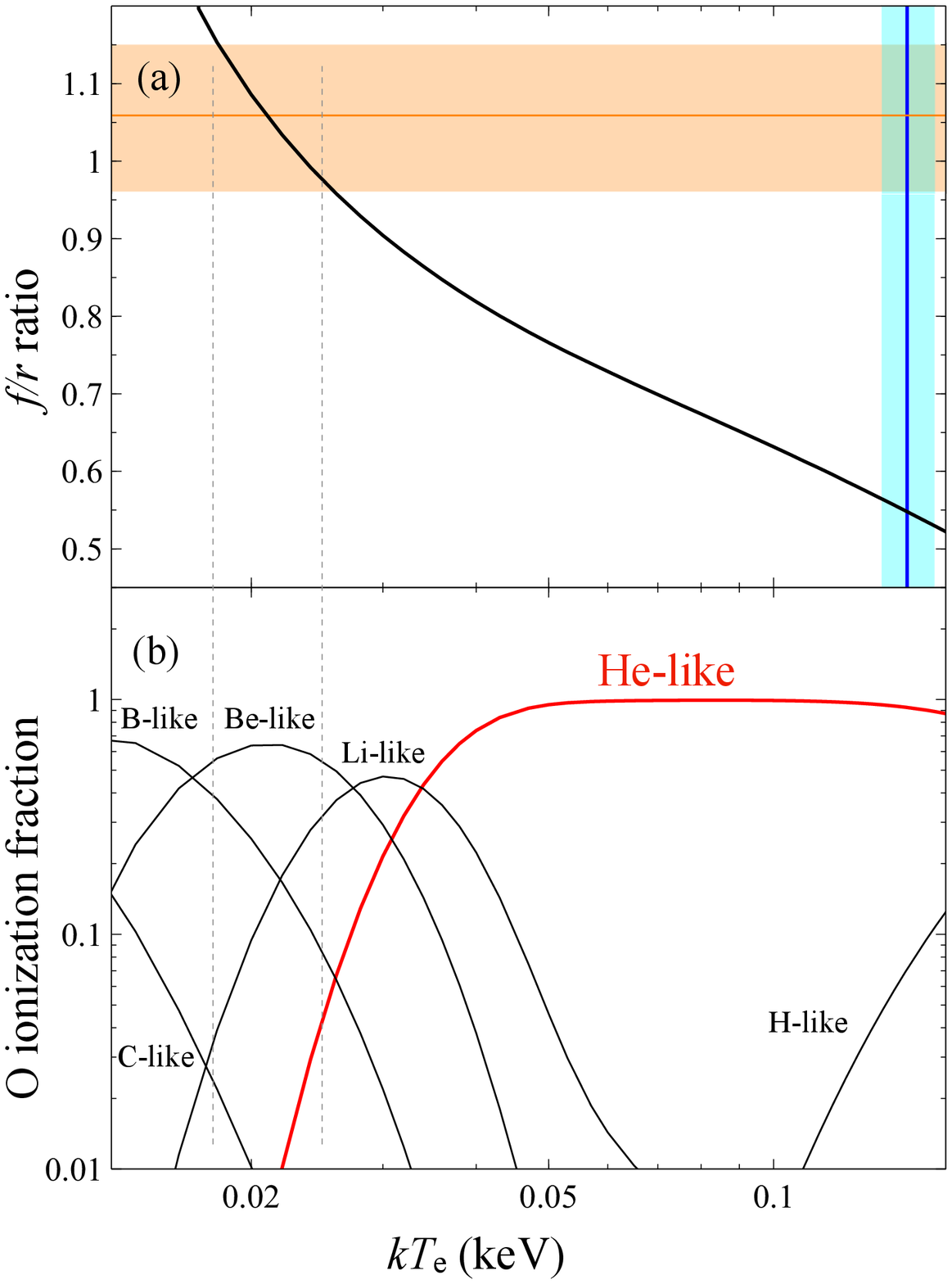}
	\end{center}
	\caption{
		(a) $f/r$ ratio of O$\emissiontype{VII}$ He$\alpha$ as a function of electron temperature $kT_{\rm{e}}$ in the case of the best-fit value of $n_et$ of the low-temperature NEI component.
		The horizontal orange and vertical blue hatched areas indicate the observed line ratio and $kT_{\rm{e}}$ of the low-temperature NEI component, respectively.
		The curve represents theoretically expected values calculated from the {\tt neij} model in SPEX.
		(b) Oxygen ionization fraction as a function of $kT_{\rm{e}}$.
	}
	\label{fig:fr_ratio}
\end{figure}

Figure~\ref{fig:neij2_spectrum} presents the MOS and RGS spectra of \J.
\citet{Mc12} reported that the X-ray spectrum of \J \ can be explained by a two-temperature non-equilibrium ionization (NEI) model (\texttt{neij}) with a power-law component for the PWN.
We first applied this ``2-NEI'' model, in which photon index $\Gamma$ and normalization of the power-law component were fixed to 2.0 and $3.5\times10^{43}~\rm{photons~s^{-1}~keV^{-1}}$ \citep{Mc12}, respectively.
Free parameters for the thermal components include the electron temperature ($kT_{\rm{e}}$), ionization timescale ($n_et$, where $n_e$ and $t$ are the electron number density and the time after shock heating, respectively), and emission measure ($n_en_{\rm{H}}V$, where $V$ is the emitting volume of the plasma).
Abundances of C, N, O, Ne, Mg, Si, and Fe were set free and tied between the two components.

The best-fit model of the 2-NEI model is plotted in Figure~\ref{fig:neij2_spectrum}, and its parameters are summarized in Table~\ref{tab:parameters}.
We found that whereas the model can reproduce the MOS spectrum, significant residuals remain in the RGS spectrum especially at the O$\emissiontype{VII}$ He$\alpha$ and O$\emissiontype{VIII}$ Ly$\beta$ lines. 
The result implies that some additional considerations are required to better explain the fine structures of the spectrum of \J.
To clarify this point, we quantified the $f/r$ intensity ratio of O$\emissiontype{VII}$ by adding four Gaussians instead of multiplet line components of O$\emissiontype{VII}$ He$\alpha$, (i.e., resonance, forbidden and intercombination lines) implemented in the \texttt{neij} code: in this method, the other lines and continua were not changed and were the same as those of the best-fit model.
We compared the obtained value with that expected from the NEI model as indicated in Figure~\ref{fig:fr_ratio} (a). 
The resultant $f/r$ ratio, $1.06^{+0.09}_{-0.10}$ requires $kT_{\rm{e}}<0.025$~keV.
On the other hand, few O$^{6+}$ (He like) ions, which emit O$\emissiontype{VII}$, are present in such a low-temperature plasma (panel (b) of Figure~\ref{fig:fr_ratio}), being inconsistent with our result.
We thus conclude that any single or multiple NEI component(s) cannot reproduce the observed RGS spectrum.
%We also rule out the possibility of an over-ionized plasma, which may explain both the $f/r$ ratio and O$\emissiontype{VIII}$ Ly$\beta$ line, since such model requires a significant excess of radiative recombination continua, which does not match the observed spectrum.
Another possible scenario to account for both $f/r$ ratio and O$\emissiontype{VIII}$ Ly$\beta$ line is an over-ionized plasma. This scenario, however, would make a significant excess of radiative recombination continua, and thus contradicts the observed spectrum. 

\begin{table*}[tp]
	\tbl{Best-fit parameters of the \J \ spectrum}{%
	\begin{tabular}{llccc}
		\hline
		Component & Parameters (unit) & 2-NEI & 2-NEI$+$CX & 2-NEI$-$Gaus (RS) \\
		\hline
		Absorption & $N_{\rm{H(Galactic)}}$ (10$^{20}$~cm$^{-2}$)& $6.0$ (fixed) & $6.0$ (fixed) & $6.0$ (fixed) \\
		 & $N_{\rm{H(LMC)}}$ (10$^{20}$~cm$^{-2}$) & $8.7^{+2.8}_{-2.7}$ & $7.2^{+2.3}_{-2.4}$ & $6.9^{+5.2}_{-3.5}$ \\
		Power law (PWN) & Normalization~($10^{44}\ \rm{photons~s^{-1}}~keV^{-1}$) & $0.35$ (fixed) & $0.35$ (fixed) & $0.35$ (fixed) \\
		 & $\Gamma$ & $2.0$ (fixed) & $2.0$ (fixed) & $2.0$ (fixed) \\
		Low-temperature NEI & Emission Measure ($10^{58}\ \rm{cm^{-3}}$) & $81^{+28}_{-20}$ & $60^{+13}_{-13}$ & $43^{+55}_{-25}$ \\
		 & $kT_{\rm{e}}$ (keV) & $0.15^{+0.01}_{-0.01}$ & $0.18^{+0.01}_{-0.02}$ & $0.15^{+0.05}_{-0.03}$ \\
		 & $n_{\rm{e}}t~(10^{11}\ \rm{cm^{-3}~s})$ & $> 10$ & $> 10$ & $> 10$ \\
		 & C & $0.34^{+0.13}_{-0.10}$ & $0.40^{+0.15}_{-0.1}$ & $0.44^{+0.47}_{-0.17}$ \\
		 & N & $0.12^{+0.05}_{-0.04}$ & $0.15^{+0.07}_{-0.04}$ & $0.18^{+0.18}_{-0.06}$ \\
		 & O & $0.26^{+0.05}_{-0.04}$ & $0.23^{+0.05}_{-0.03}$ & $0.53^{+0.39}_{-0.10}$ \\
		 & Ne & $0.34^{+0.07}_{-0.05}$ & $0.33^{+0.08}_{-0.06}$ & $0.38^{+0.31}_{-0.07}$ \\
		 & Mg & $0.42^{+0.08}_{-0.07}$ & $0.41^{+0.13}_{-0.08}$ & $0.51^{+0.27}_{-0.09}$ \\
		 & Si & $0.22^{+0.09}_{-0.08}$ & $0.19^{+0.11}_{-0.07}$ & $0.32^{+0.20}_{-0.11}$ \\
		 & Fe & $0.25^{+0.04}_{-0.03}$ & $0.22^{+0.06}_{-0.03}$ & $0.25^{+0.06}_{-0.04}$ \\
		High-temperature NEI & Emission Measure ($10^{58}\ \rm{cm^{-3}}$) & $12^{+5}_{-4}$ & $7.9^{+7.9}_{-3.5}$ & $17^{+10}_{-16}$ \\
		 & $kT_{\rm{e}}$ (keV) & $0.42^{+0.05}_{-0.03}$ & $0.48^{+0.09}_{-0.08}$ & $0.35^{+0.33}_{-0.08}$ \\
		 & $n_{\rm{e}}t~(10^{11}\ \rm{cm^{-3}~s})$ & $1.6^{+0.5}_{-0.3}$ & $1.4^{+0.5}_{-0.3}$ & $> 2.2$ \\
		CX & Emission Measure ($10^{58}\ \rm{cm^{-3}}$) & $\cdots$ & $18^{+100}_{-9}$ & $\cdots$ \\
		 & $v_{\rm{col}}~(\rm{km~s^{-1}})$ & $\cdots$ & $< 286$ & $\cdots$ \\
		Negative Gaussian\footnotemark[$*$]: Ne$\emissiontype{IX}$ He$\alpha$ ($r$)  & Normalization~($10^{44}\ \rm{photons~s^{-1}}$) & $\cdots$  & $\cdots$ & $<8.8\times10^{-2}$ \\
		~~~~~~~~~~~~~~~~~~~~~~~~~~~~Fe$\emissiontype{XVII}$ L(3d-2p) &  Normalization~($10^{44}\ \rm{photons~s^{-1}}$) & $\cdots$ & $\cdots$ & $0.14\pm{0.09}$ \\
		~~~~~~~~~~~~~~~~~~~~~~~~~~~~Fe$\emissiontype{XVII}$ L(3s-2p) &  Normalization~($10^{44}\ \rm{photons~s^{-1}}$) & $\cdots$ & $\cdots$ & $<0.31$ \\
		~~~~~~~~~~~~~~~~~~~~~~~~~~~~O$\emissiontype{VIII}$ Ly$\alpha$ &  Normalization~($10^{44}\ \rm{photons~s^{-1}}$) & $\cdots$ & $\cdots$ & $2.3^{+1.2}_{-0.6}$ \\
		~~~~~~~~~~~~~~~~~~~~~~~~~~~~O$\emissiontype{VII}$ He$\alpha$ ($r$) &  Normalization~($10^{44}\ \rm{photons~s^{-1}}$) & $\cdots$ & $\cdots$ & $1.9^{+0.6}_{-0.4}$ \\
		\hline
		 & W-statistic/d.o.f. & 4124/3627 & 4107/3625 & 4085/3622 \\
		\hline
	\end{tabular}}\label{tab:parameters}
	\begin{tabnote}
		\footnotemark[$*$]
		The line centroid wavelengths of the Gaussians at Ne$\emissiontype{IX}$ He$\alpha$ ($r$), 
		Fe$\emissiontype{XVII}$ L(3d-2p), Fe$\emissiontype{XVII}$ L(3s-2p), O$\emissiontype{VIII}$ Ly, and O$\emissiontype{VII}$ He$\alpha$ ($r$)
		are fixed to 13.4~\AA, 15.0~\AA, 17.0~\AA, 18.9~\AA, and 21.6~\AA, respectively.
		\end{tabnote}
\end{table*}

\begin{figure*}[ht]
	\begin{center}
	\includegraphics[width=90mm]{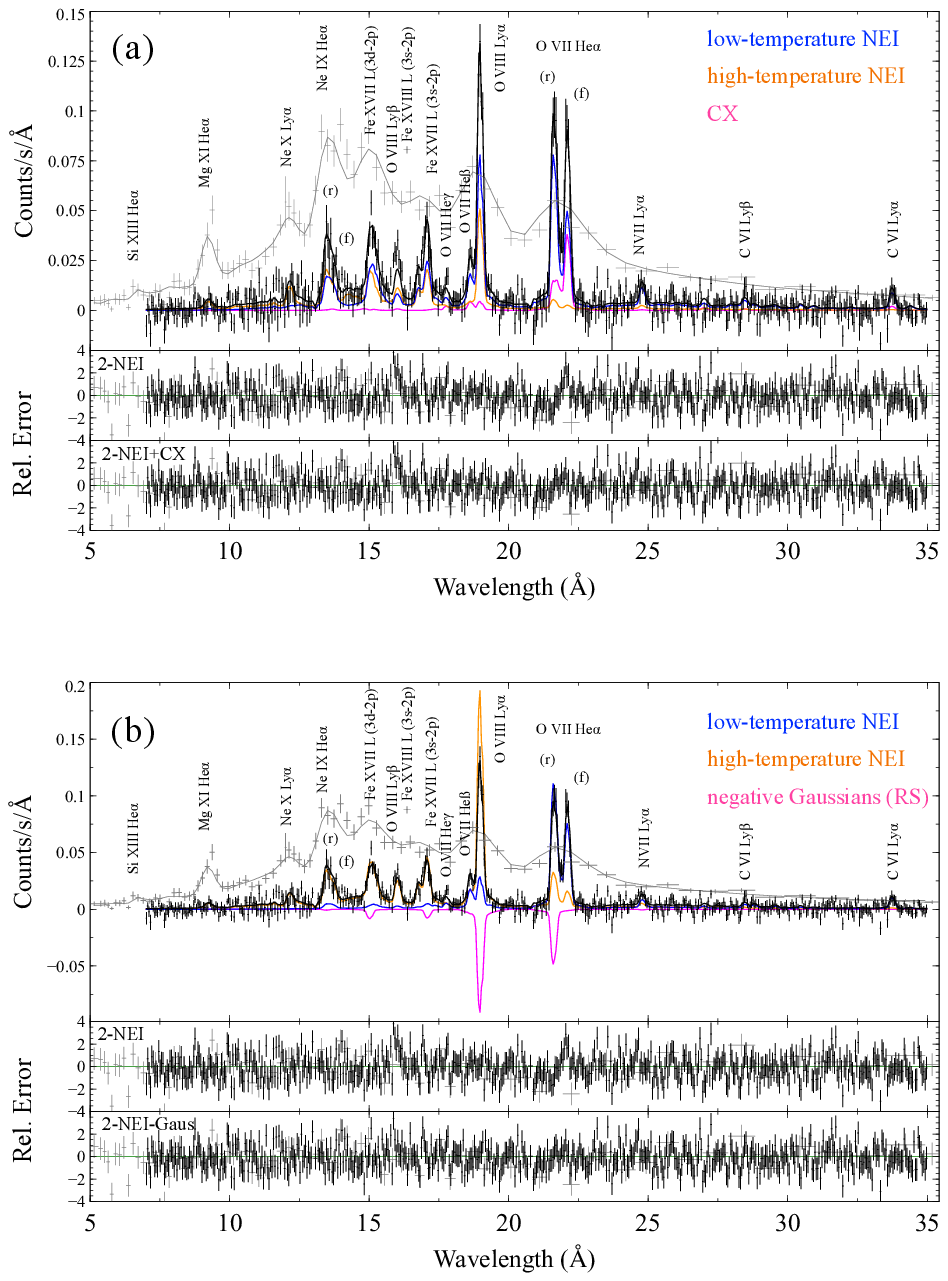}
	\end{center}
	\caption{
	(a)Same as Figure~\ref{fig:neij2_spectrum} but with the 2-NEI$+$CX model.
	The magenta curve represents the CX component.
	The middle and lower panels show residuals from the 2-NEI and 2-NEI$+$CX models, respectively.
	(b)Same as Figure~\ref{fig:neij2_spectrum} but with the 2-NEI$-$Gaus model.
	The magenta curve indicates scattered line intensities.
	The middle and lower panels show residuals from the 2NEI and 2-NEI$-$Gaus models, respectively.
	}
	\label{fig:cx_spectrum}
\end{figure*}

We next added a CX component to the 2-NEI model (hereafter, 2-NEI$+$CX model) to enhance the forbidden line intensity of O$\emissiontype{VII}$, as \citet{Uc19} did for a similar case of the Cygnus Loop.
Free parameters of the CX model are normalization ($n_{\rm{H}}n_{nh}V$, where $n_{nh}$ is the density of the neutral materials) and shock velocity ($zv$).
The ionization temperature was tied to $kT_{\rm{e}}$ of the low-temperature NEI component.
The best-fit result and parameters  are shown in Figure~\ref{fig:cx_spectrum} and Table~\ref{tab:parameters}, respectively.
As a result, we successfully fitted the RGS spectrum with the 2-NEI$+$CX model, except that the discrepancy between the data and model is still seen at the O$\emissiontype{VIII}$ Ly$\beta$ line.

A similar excess of the O$\emissiontype{VIII}$ Ly$\beta$ line has been reported by \citet{Am20}, who analyzed the RGS spectrum of N49 and concluded that the line ratios including $f/r$ of O$\emissiontype{VII}$ can be reasonably explained by  taking into account the effect of RS. 
We thus applied negative Gaussians in addition to the 2-NEI model (hereafter, 2-NEI$-$Gaus model), in which we assumed that the SNR shell is a slab and all scattered photons will escape from the line of sight \citep{Ka95}.
The Gaussians were fixed at the centroid wavelengths of the lines whose oscillator strengths are relatively large: resonance lines of Ne$\emissiontype{IX}$ and O$\emissiontype{VII}$, Fe$\emissiontype{XVII}$ L(3d-2p), Fe$\emissiontype{XVII}$ L(3s-2p), and O$\emissiontype{VIII}$ Ly$\alpha$.
Normalizations (photons $\rm{s^{-1}}$) of these five Gaussians were set free and the other parameters are the same as the 2-NEI model.
As shown in Figure~\ref{fig:cx_spectrum}, the 2-NEI$-$Gaus model globally reduces the residuals.
The best-fit parameters (Table~\ref{tab:parameters}) are consistent with those expected for a typical middle-aged SNR.
We therefore claim that a presence of RS cannot be ruled out in terms of the spectral fitting.

\section{Discussion}\label{sec:disc}
As indicated in the previous section, the high-resolution X-ray spectrum of  \J \ suggests the presence of CX or RS in the remnant.
Similar cases have often been discussed in the literature \citep[e.g.,][]{Uc19, Am20, Su20}.
Although it is in general difficult to distinguish between these two possibilities with the available spectroscopies, the SNR morphology and surrounding environment may provide a clue to the true origin of the high $f/r$ ratio.
In Table \ref{tab:f/r}, we summarize LMC/SMC SNRs for which the $f/r$ (or $G$) ratios were measured so far using the RGS to compare our results with those from other SNRs in the discussion below.

\begin{figure}[ht]
	\begin{center}
	\includegraphics[width=0.35\textwidth]{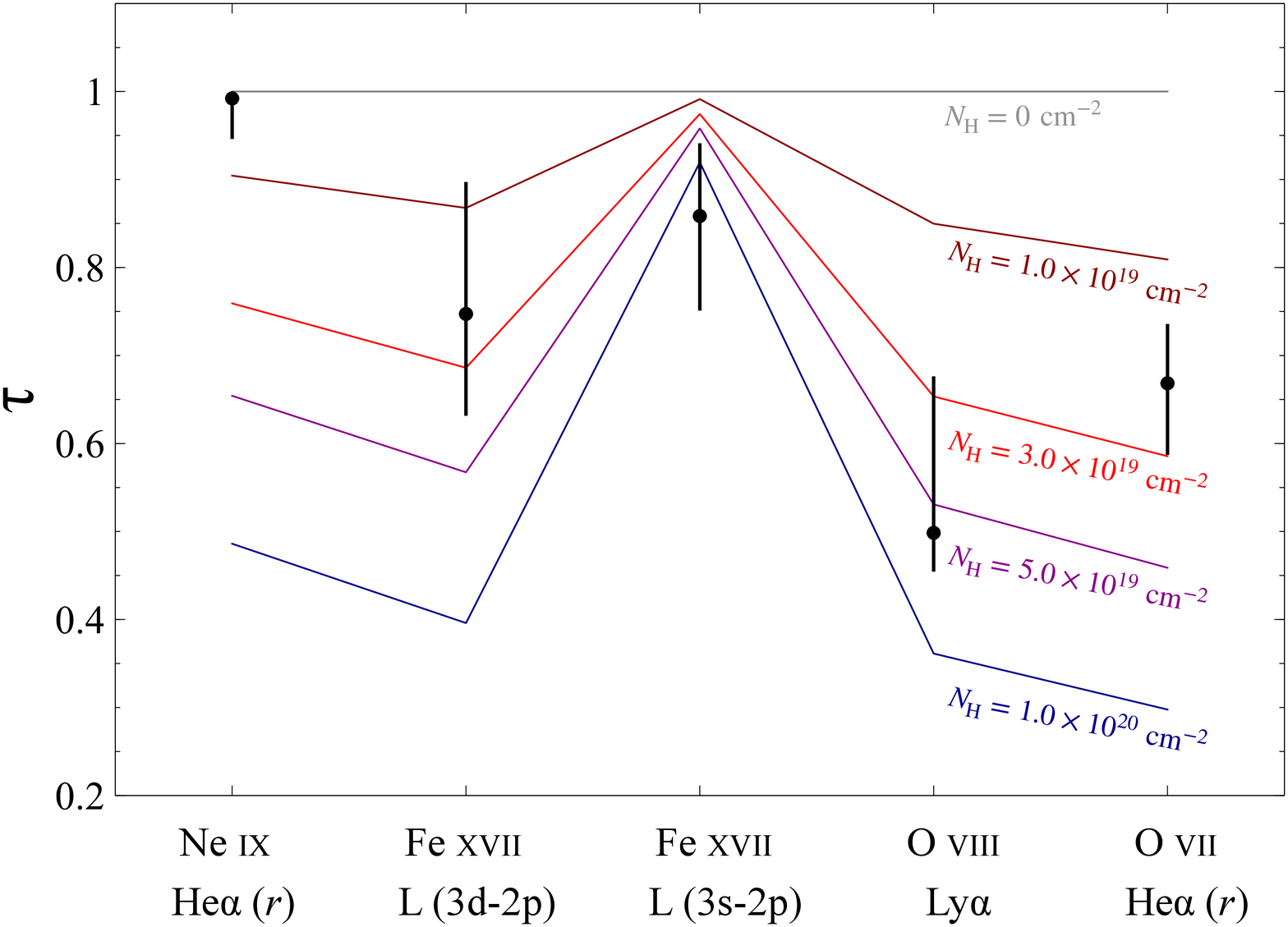}
	\end{center}
	\caption{
	Transmission factors $\tau$ for each line.
	The black points are those estimated from the observed line intensities and normalizations of the negative Gaussians.
	The colored curves represent expected values of $\tau$ as a function of $N_{\rm{H}}$.
	}
	\label{fig:rs_plot}
\end{figure}

\begin{figure}[ht]
	\begin{center}
	\includegraphics[width=0.4\textwidth]{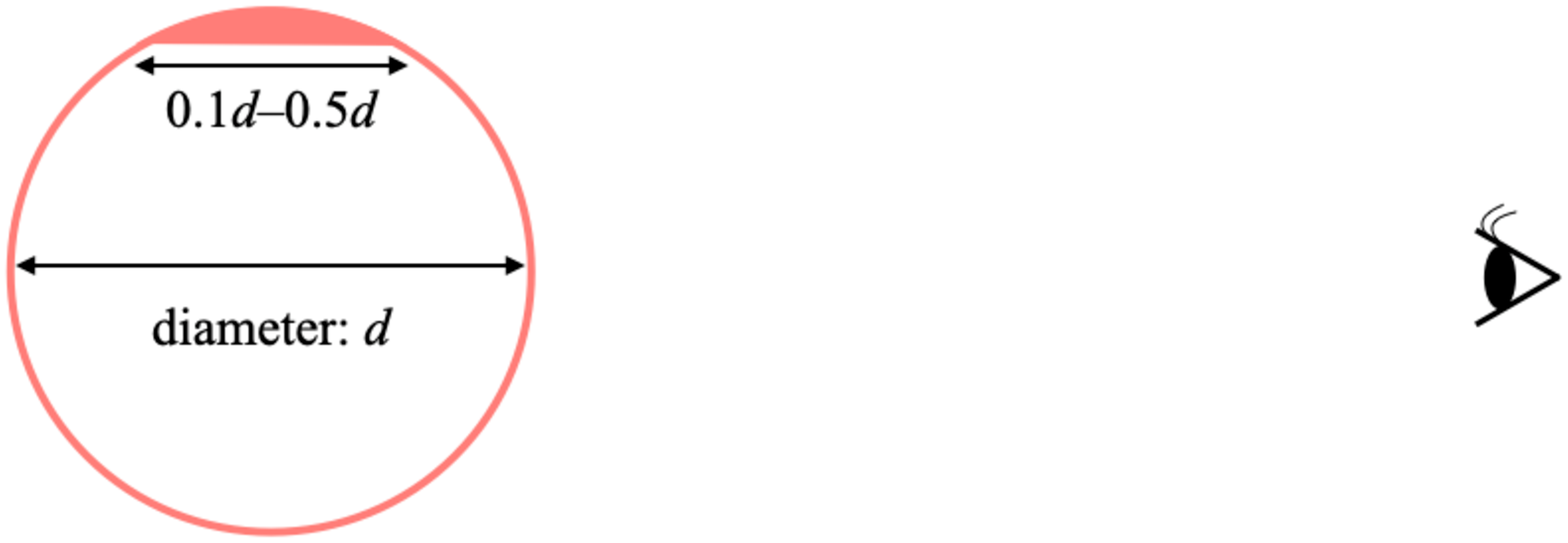}
	\end{center}
	\caption{
	Example schematic view of an X-ray emitting plasma that accounts for the result of the calculation of RS. 
	}
	\label{fig:RS_picture}
\end{figure}

\subsection{X-ray Morphology of \J}
For quantitative evaluation of the effect of RS, we calculated a transmission factor $\tau$ by applying the same method as \citet{Am20}.
We compared theoretical values of $\tau$ for several optical depths according to \citet{Ka95} with those estimated from the best-fit normalizations of the negative Gaussians.
Figure~\ref{fig:rs_plot} shows the result.
Although the model fails to explain the observed resonance line of Ne$\emissiontype{IX}$, the estimated $\tau$ for \J \ are roughly consistent with those at $N_{\rm{H}}=1.0$--$5.0\times10^{19}$~cm$^{-2}$, which corresponds to a plasma depth of 3--18~pc under an assumption of the plasma density $n_{\rm{H}}=1.1$~cm$^{-3}$ \citep{Wi06}.
Since the diameter of \J \ is estimated to be $\sim36$~pc from the apparent angular size of the shell ($\sim2.5'$), the line-of-sight plasma depth that contributes to RS should be 10--50\% of the diameter.
If this is really the case, the shell-type SNR would be required to have a highly asymmetric morphology; for instance, a bright shell is prominent only on one side of the remnant (see Figure~\ref{fig:RS_picture}).

If \J \ has an ideal spherical symmetric structure, the RS effect will be cancelled out and an enhancement of $f/r$ will not occur.
Note that the RGS spectrum of \J \ was obtained from the entire region.
We can thus postulate that asymmetry of an SNR is a key parameter to evaluate the effect of RS.
From a soft-band imaging analysis by \citet{Lo11}, we found that \J \ has a less asymmetric morphology among the six SNRs listed in Table~\ref{tab:f/r}.
Other core-collapse remnants are more ``elliptical'' (N23) or ``non-uniform'' (N132D), which are parametrized as $P_2/P_0$ and $P_3/P_0$ in their calculation.
Although N49 was not analyzed by \citet{Lo11}, it would also be categorized as a highly elliptical remnant due to its morphology similar to that of N23.
It is reasonable that N23, N49, and N132D show relatively high $f/r$ ratios due to RS, as claimed by previous studies \citep{Br11, Am20, Su20}.
On the other hand, in the case of \J, the effect of RS might be unlikely or insufficient to satisfactorily explain the observed high $f/r$ ratio.

\begin{table*}[hp]
	\tbl{$f/r$ ratios or $G$-ratios for O$\emissiontype{VII}$ He$\alpha$}{%
	\begin{tabular}{lccrclccc}
		\hline
		Name & Type of SNe & Ref. & Age & Ref. & f/r or G ratios & Ref. & Surrounding & Ref. Surrounding \\
		 & & Type\footnotemark[$*$] & (yr) & Age\footnotemark[$*$] & & Ratios\footnotemark[$*$] & environments & environments\footnotemark[$*$] \\
		\hline
		1E~0102$-$7219 & Ib/c or IIL/b & 1, 2 & 1000 & 3 & $0.55$\footnotemark[$\ddag$]$\pm0.03$ & 4 & no data &  \\
		N132D & Ib & 1 & 2500 & 5 & $0.68$\footnotemark[$\dag$]$\pm0.02$ & 6 & CO and H\emissiontype{I} clouds & 7, 8, 9 \\
  		DEM~L71 & Ia & 10 & 4400 & 11 & $0.65$\footnotemark[$\dag$] & 12 & no data &  \\
        N23 (0506$-$68.0) & II & 13 & 4600 & 13 & $0.99$\footnotemark[$\ddag$]$\pm0.06$ & 14 & star-forming region & 15 \\
		N49 & II & 16 & 6600 & 17 & $1.23$\footnotemark[$\ddag$]$\pm0.12$ & 18 & CO and H\emissiontype{I} clouds & 7, 8, 19 \\
		\J & II & 20 & 13000 & 21 & $1.06$\footnotemark[$\dag$]$^{+0.09}_{-0.10}$ & 22 & H\emissiontype{I} clouds & 22 \\
		\hline
	\end{tabular}}\label{tab:f/r}
	\begin{tabnote}
	\footnotemark[$*$]
	References.
	(1) \cite{Bl00};
	(2) \cite{Ch05};
	(3) \cite{Hu00};
	(4) \cite{Ra01};
	(5) \cite{Vo11};
	(6) \cite{Su20};
	(7) \cite{Ba97};
	(8) \cite{Sa17};
	(9) \cite{Sa20};
	(10) \cite{Hu98};
	(11) \cite{Gh03};
	(12) \cite{He03};
	(13) \cite{Hu06};
	(14) \cite{Br11};
	(15) \cite{Ch88};
	(16) \cite{Uc15};
	(17) \cite{Pa03};
	(18) \cite{Am20};
	(19) \cite{Ya18};
	(20) \cite{Lo09};
	(21) \cite{Ga03};
	(22) This work.
	\\
	\footnotemark[$\dag$]
	$f/r$ ratio
	\footnotemark[$\ddag$]
	$G$ ratio $(f+i)/r$ \\
	\end{tabnote}
\end{table*}

\subsection{Surrounding Environment of \J}

\begin{figure*}[th]
	\begin{center}
		\includegraphics[width=130mm]{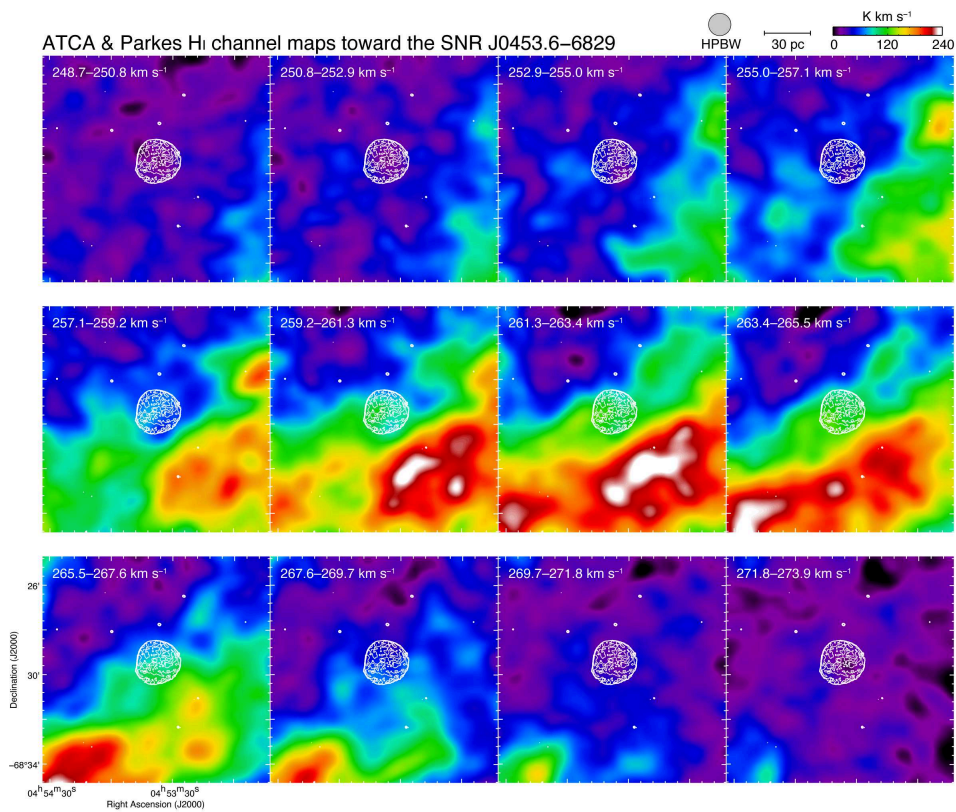}
	\end{center}
	\caption{
	ATCA \& Parkes H\emissiontype{I} channel maps overlaid with the Chandra X-ray intensity of \J \ (white contours).
	Each panel shows the H\emissiontype{I} intensity map integrated over the 2.1~$\rm{km~s^{-1}}$ width evenly spaced in the 248.7--273.9~$\rm{km~s^{-1}}$ range.
	}
	\label{fig:HI_channel}
\end{figure*}

\begin{figure*}[htp]
	\begin{center}
		\includegraphics[width=100mm]{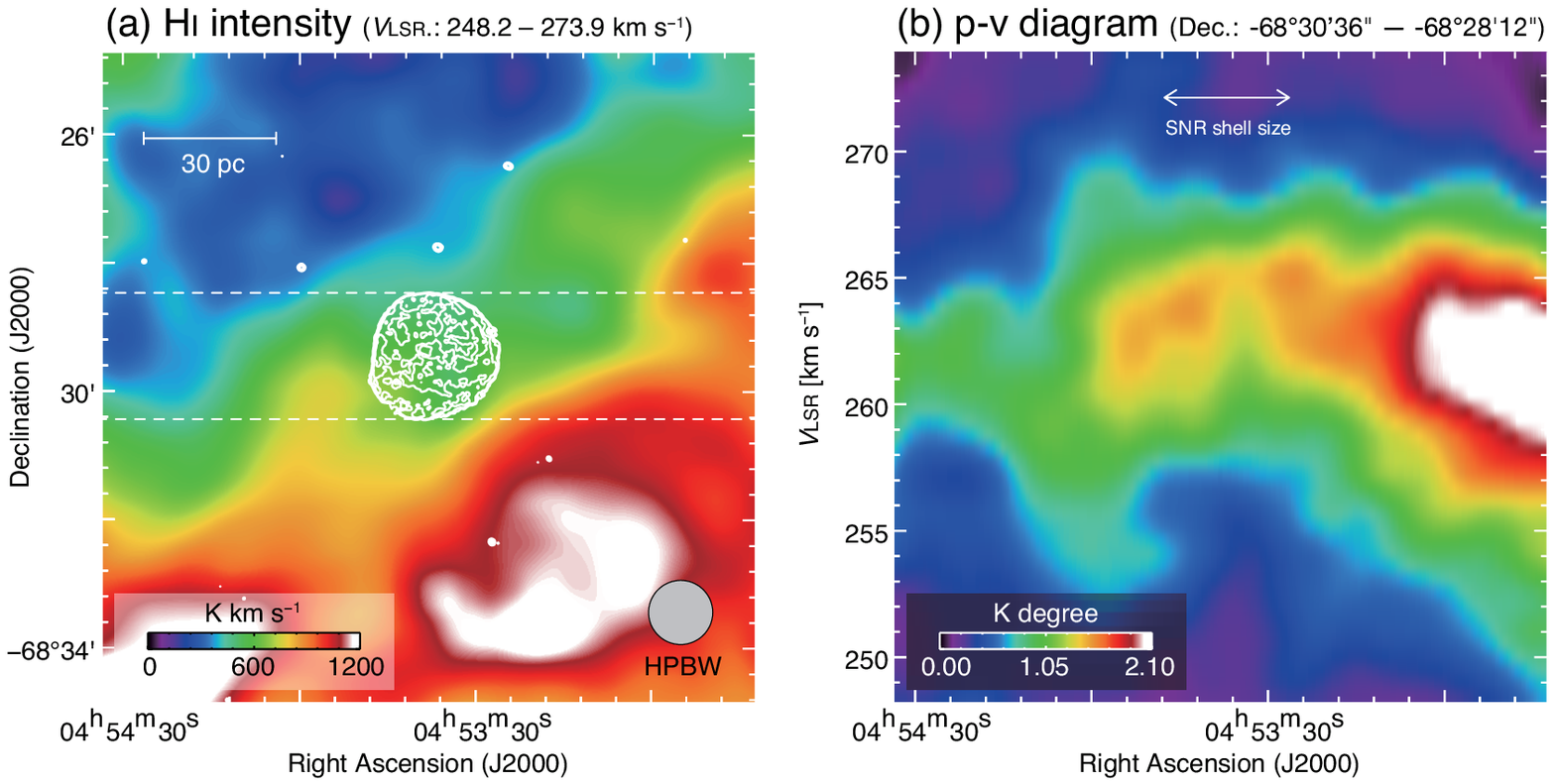}
	\end{center}
	\caption{
	(a) Integrated intensity map of the ATCA \& Parkes H\emissiontype{I} in $V_{\rm{LSR}}$ of 248.2--273.9~$\rm{km~s^{-1}}$.
	(b) Position-velocity diagram of H\emissiontype{I} image.
	The integration range in Dec. is from \timeform{-68D30'36''} to \timeform{-68D28'12''} (J2000.0).
	The white arrow indicates the position of \J.
	}
	\label{fig:HI_map}
\end{figure*}

CX is another possibility that causes the enhancement of $f/r$.
An interaction with a dense ambient medium is expected in this case, as in previous studies of Galactic SNRs with the RGS: Puppis~A \citep{Ka12} and the Cygnus Loop \citep{Uc19}.
While \citet{Mc12} implied a presence of dense gas in the vicinity of \J \ because of a spatial correlation between the X-ray and infrared morphologies, the surrounding environment of this remnant has still been unclear \citep[][]{Wi06, La15}.
As shown in Figure \ref{fig:HI_channel}, we compared the ATCA \& Parkes \citep{Ki03} H\emissiontype{I} velocity channnel map around \J \  with the X-ray morphology and found H\emissiontype{I} clouds located along with the southwestern half of \J.
Figure~\ref{fig:HI_map} shows the integrated intensity maps of H\emissiontype{I}.
We also found the southwestern part of the remnant is increasingly covered with an H\emissiontype{I} cloud.
The position-velocity diagram  suggests that the SNR shell is expanding into the dense gas (panel (b) of Figure~\ref{fig:HI_map}).

%If the SNR shell interacts with the H\emissiontype{I} cloud in the southwestern region, relatively strong forbidden line emission would be detected there.
%We thus compared monochromatic images of resonance and forbidden lines of O$\emissiontype{VII}$ (Figure~\ref{fig:rgs_img}) using a reconstruction method given by \citet{He03}.
%As a result, the forbidden line is most prominent in the west, which is roughly consistent with the above expectation.
%The emitting region is, however, concentrated in the north of \J.
%This may be in conflict with a standard picture drawn by \citet{La04}, in which CX is expected only at an outermost rim ($<1\%$ of the radius) of an SNR.
%One interpretation of this is that a line-of-sight bright shell corresponds to the shock interaction region where the CX dominantly occurs; the bright ring-like structure in the west \citep{Ha12, Sc16} is indeed the region where the forbidden line is enhanced (panels~(a) and (d) of Figure~\ref{fig:rgs_img}), which may support our idea.

\begin{figure}[ht]
	\begin{center}
	\includegraphics[width=0.5\textwidth]{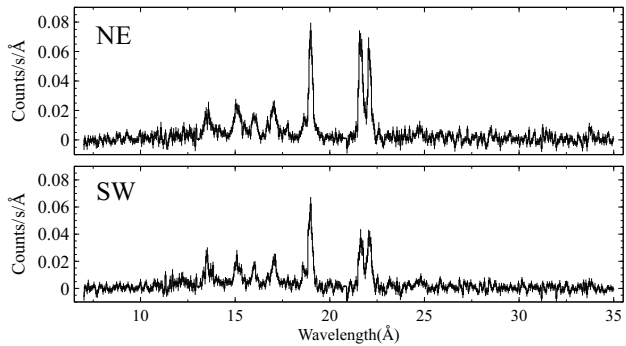}
	\end{center}
	\caption{
	RGS1$+$2 spectra of the NE (top) and SW (bottom) regions of \J.
		}
	\label{fig:NESW}
\end{figure}

\begin{figure*}[th]
	\begin{center}
	\includegraphics[width=90mm]{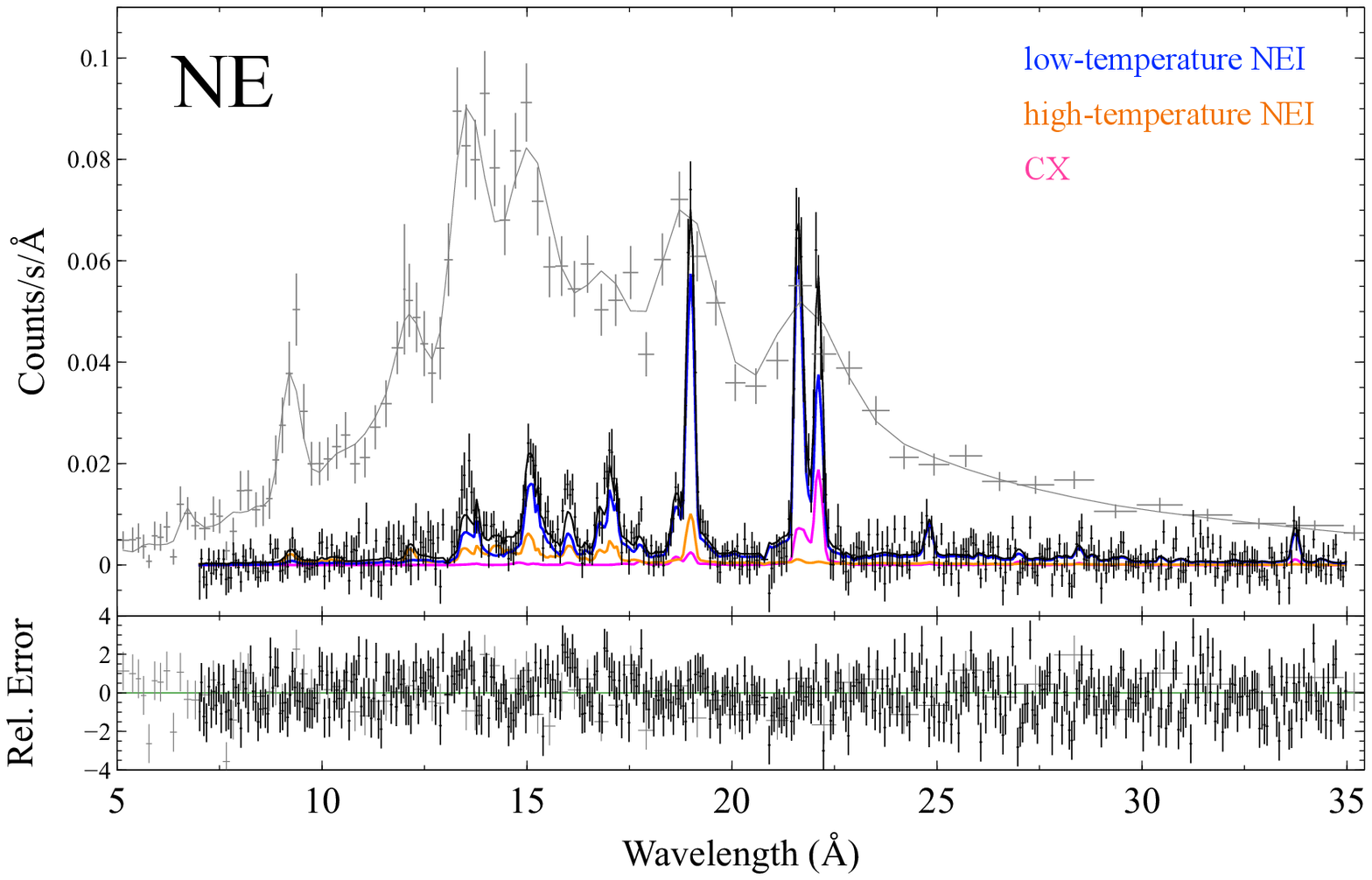}
	\includegraphics[width=90mm]{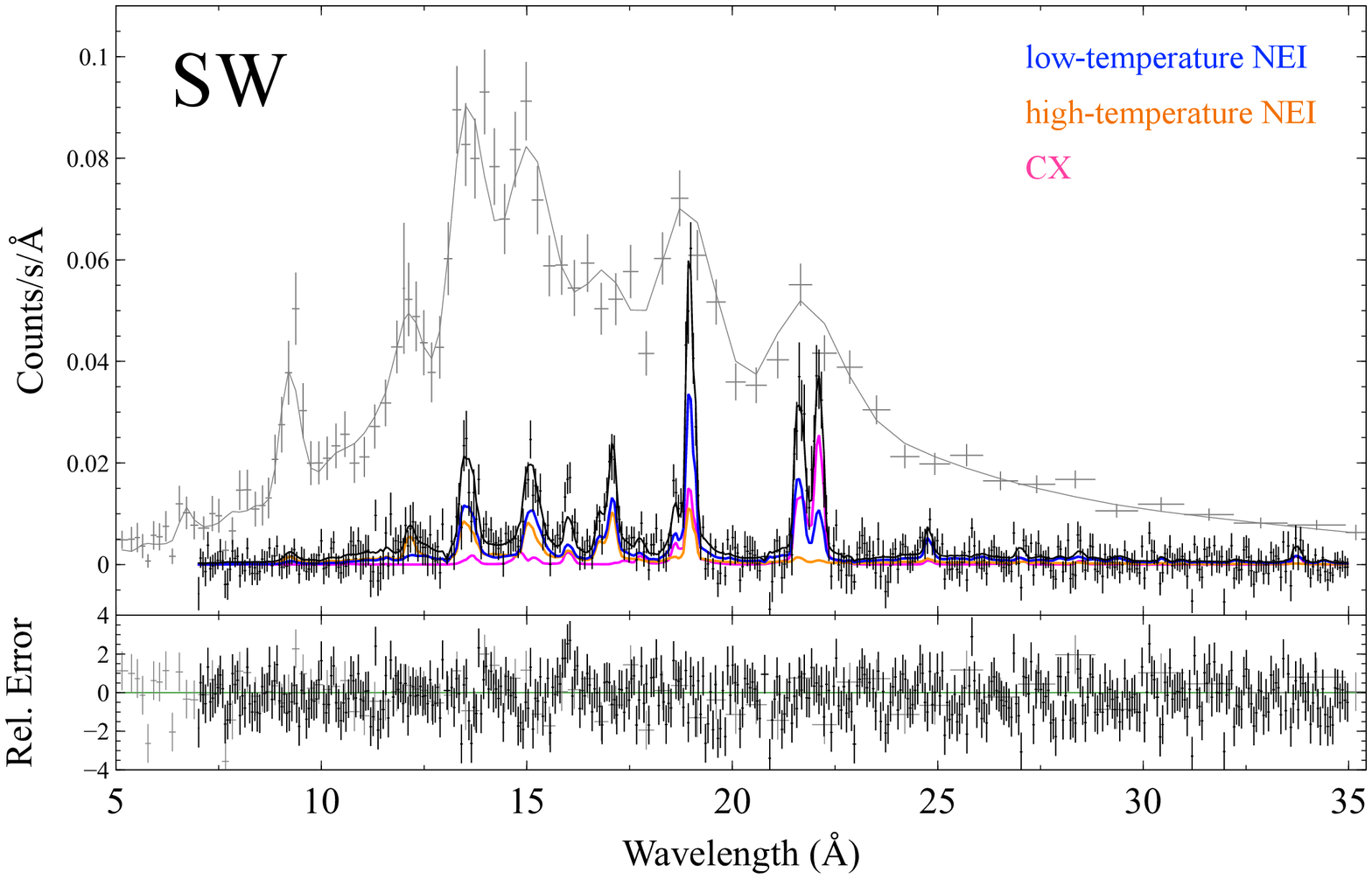}
	\end{center}
	\caption{
	(a)Same as the panel~(a) of Figure~\ref{fig:neij2_spectrum} but for the NE (top) and SW (bottom) regions.
	}
	\label{fig:cx_NESW}
\end{figure*}

If the SNR shell interacts with the H\emissiontype{I} cloud in the southwestern region, relatively strong forbidden line emission would be detected there.
We thus divided the data into two in the cross-dispersion direction (namely, northeast; NE and southwest; SW) as indicated in Figure~\ref{fig:full_img}.
As displayed in Figure~\ref{fig:NESW}, the forbidden line intensity of O$\emissiontype{VII}$ in SW is stronger than that in NE.
Applying the same method as the entire region, we obtained $f/r$ ratios of $0.97^{+0.18}_{-0.14}$ and $1.41^{+0.40}_{-0.29}$  for NE and SW, respectively.
Although statistically they are equal  within the errors, the trend is consistent with the above expectation and thus strongly supports the presence of CX.
The best-fit models of 2-NEI$+$CX for these regions are displayed in Figure~\ref{fig:cx_NESW}. 
The model parameters  are given in Table~\ref{tab:parameters_NESW}.
While the CX component is required both in NE and SW, its contribution is relatively dominant in SW.
We thus confirmed that the anomalous $f/r$ ratios is due to the CX emission, mainly caused by an  interaction with the southwestern H\emissiontype{I} cloud.

According to the discussion above, we presume that the emitting region of CX is the southwestern edge of \J, which is in contact with the H\emissiontype{I} cloud.
%$<1\%$ of the SNR radius along the line of sight of \J.
To quantitatively examine the  possibility of CX,  we estimate the emitting volume $V_{\rm CX}$ using the volume emission measure of the CX component ${\rm EM_{CX}}=n_{\rm{H}} n_{\rm{nh}} V_{\rm CX}$, where $n_{\rm{nh}}$ is  the neutral material density of the surrounding gas.
Given that the H\emissiontype{I} gas has a typical density of $n_{\rm{nh}}=10$~cm$^{-3}$, we obtain the emitting volume $V_{\rm CX}$ to be $2\times10^{58}$~cm${^3}$.
Since the total volume of \J \ is estimated to be $V_{\rm SNR}\sim10^{60}$~cm${^3}$ assuming a diameter of $\sim36$~pc, we conclude that the CX occurs in $\sim0.4\%$ of the SNR radius.
The result fits well with the calculation by \citet{La04} and thus supports the possibility that the observed anomalous $f/r$ ratio is due to CX.
Note that the significant residuals seen at $\sim$16~\AA \ (section~\ref{sec:analysis}) is still an open question; such discrepancies around the O$\emissiontype{VIII}$ Ly$\beta$ line are often pointed out by many RGS observations \citep[e.g.,][]{Am20}, and might be due to uncertainties in the atomic data \citep[see also,][]{de12}.

%\begin{figure*}[h]
%	\begin{center}
%	\includegraphics[width=\textwidth]{rgs_img.eps}
%	\end{center}
%	\caption{
%	(a) Combined EPIC (MOS and pn) image of \J \ in the 0.3--8.0 keV band.
%	The white and magenta circles indicate a rough position of the shell and the bright ring-like structure (see text), respectively.
%	(b) RGS monochromatic image of the resonance line of O$\emissiontype{VII}$ ($r$).
%	(c) Same as (b) but for the forbidden line of O$\emissiontype{VII}$ ($f$).
%	(d) Two-color monochromatic image of $r$ (red) and $f$ (green).
%	}
%	\label{fig:rgs_img}
%\end{figure*}

\section{Conclusions}\label{sec:conc}
We performed a high-resolution spectroscopy of \J \ with the RGS onboard XMM-Newton and found that the intensity of the forbidden line of O$\emissiontype{VII}$ is significantly stronger than expected from a simple thermal (2-NEI) model.
To account for the obtained high $f/r$ ratio ($1.06^{+0.09}_{-0.10}$), we examined two possibilities: CX and RS, which have been proposed for explaining similar spectral features found in SNRs.
Both models are statistically acceptable, although small residuals remain at $\sim$16~\AA \ (around the O$\emissiontype{VIII}$ Ly$\beta$ line) between the data and the 2-NEI$+$CX model.
Such discrepancies are often pointed out by many RGS observations \citep[e.g.,][]{de12, Am20} and are likely due to uncertainties in the atomic data.
From the best-fit result with the RS model, we estimated a transmission factor $\tau$; the result requires a significantly asymmetric  shape along the line of sight.
This may be inconsistent with the apparent morphology of \J, since a  previous systematic X-ray study indicates that this remnant is one of the ``least asymmetric'' core-collapse SNRs \citep{Lo11}.
On the other hand, our estimate of the emitting volume for the CX component ($\sim0.4\%$ of the SNR radius) agrees well with a theoretical expectation \citep{La04}.
We also found evidence of an interaction between \J \ and the dense ambient gas in the  ATCA \& Parkes H\emissiontype{I} map, which supports the picture that the observed $f/r$ ratio is due to the CX emission at SNR shock fronts.
In conclusion, the presence of CX in \J \ is favored in our study, while a slight or significant contribution of the RS effect also cannot be ruled out.
Future spatially resolved spectroscopies with high angular resolution missions like Athena will clarify this point.

\begin{table*}[p]
	\tbl{Best-fit parameters of the NE and SW spectra}{%
	\begin{tabular}{llcc}
		\hline
		Component & Parameters (unit) &  \multicolumn{2}{c}{2-NEI$+$CX}  \\
		\hline
		 &  &  NE & SW \\
		Absorption & $N_{\rm{H(Galactic)}}$ (10$^{20}$~cm$^{-2}$)&  \multicolumn{2}{c}{$6.0$ (fixed)}  \\
		 & $N_{\rm{H(LMC)}}$ (10$^{20}$~cm$^{-2}$) & $7.0^{+2.3}_{-3.4}$ & $6.3^{+3.3}_{-2.5}$   \\
		Power law (PWN) & Normalization~($10^{44}\ \rm{photons~s^{-1}}~keV^{-1}$) &  \multicolumn{2}{c}{$0.35$ (fixed)}   \\
		 & $\Gamma$ &  \multicolumn{2}{c}{$2.0$ (fixed)}    \\
		Low-temperature NEI & Emission Measure ($10^{58}\ \rm{cm^{-3}}$) & $100^{+25}_{-48}$ & $33^{+14}_{-17}$  \\
		 & $kT_{\rm{e}}$ (keV) & $0.20^{+0.04}_{-0.01}$ & $0.21\pm0.01$ \\
		 & $n_{\rm{e}}t~(10^{11}\ \rm{cm^{-3}~s})$ & $>2$ & $>0.6$ \\
		 & C & $0.6\pm0.2$ & $0.2\pm0.1$ \\
		 & N & $0.3\pm0.1$ &$0.17^{+0.06}_{-0.07}$  \\
		 & O & $0.27^{+0.06}_{-0.07}$ &$0.17^{+0.04}_{-0.05}$   \\
		 & Ne & $0.4\pm0.1$ & $0.29^{+0.06}_{-0.08}$   \\
		 & Mg & $0.5\pm0.1$ & $0.39^{+0.08}_{-0.09}$   \\
		 & Si & $0.3\pm0.1$ & $0.2^{+0.2}_{-0.1}$   \\
		 & Fe & $0.24^{+0.06}_{-0.05}$ & $0.16\pm0.04$   \\
		High-temperature NEI & Emission Measure ($10^{58}\ \rm{cm^{-3}}$) & $8^{+6}_{-2}$ & $8^{+21}_{-3}$   \\
		 & $kT_{\rm{e}}$ (keV) & $0.59^{+0.09}_{-0.10}$ & $0.37^{+0.07}_{-0.09}$  \\
		 & $n_{\rm{e}}t~(10^{11}\ \rm{cm^{-3}~s})$ & $1.7^{+3.6}_{-0.6}$  & $>3$ \\
		CX & Emission Measure ($10^{58}\ \rm{cm^{-3}}$) & $16^{+75}_{-14}$ & $20^{+27}_{-9}$  \\
		 & $v_{\rm{col}}~(\rm{km~s^{-1}})$ & $< 250$ & $350^{+180}_{-130}$  \\
		\hline
		 & W-statistic/d.o.f. & 4108/3625 & 4107/3625 \\
		\hline
	\end{tabular}}\label{tab:parameters_NESW}
\end{table*}

\begin{ack}
	We thank Brian J. Williams for a helpful discussion about the previous multiwavelength studies of the SNR J0453.6$-$6829.
	The ATCA and the Parkes radio telescope are all part of the Australia Telescope National Facility, which is funded by the Australian Government for operation as a National Facility managed by CSIRO. We acknowledge the Gomeroi and Wiradjuri people as the traditional owners of the Observatory sites.
	This work is supported by JSPS/MEXT KAKENHI Scientific Research Grant Numbers JP19K03915 (H.U.), JP19H01936 (T.T.), JP19K14758 (H.S.), JP20KK0309 (H.S.), and JP21H04493 (T.G.T. and T.T.).
\end{ack}


\begin{thebibliography}{}
	\bibitem[Amano et al.(2020)]{Am20} Amano, Y., Uchida, H., Tanaka, T., et al.\ 2020, \apj, 897, 12. 
	\bibitem[Banas et al.(1997)]{Ba97} Banas, K.~R., Hughes, J.~P., Bronfman, L., et al.\ 1997, \apj, 480, 607.
	\bibitem[Bhardwaj et al.(2007)]{Bh07} Bhardwaj, A., Elsner, R.~F., Randall Gladstone, G., et al.\ 2007, \planss, 55, 1135. 
	\bibitem[Blair et al.(2000)]{Bl00} Blair, W.~P., Morse, J.~A., Raymond, J.~C., et al.\ 2000, \apj, 537, 667.
	\bibitem[Broersen et al.(2011)]{Br11} Broersen, S., Vink, J., Kaastra, J., et al.\ 2011, \aap, 535, A11. 
	\bibitem[Chevalier(2005)]{Ch05} Chevalier, R.~A.\ 2005, \apj, 619, 839.
	\bibitem[Chu \& Kennicutt(1988)]{Ch88} Chu, Y.-H. \& Kennicutt, R.~C.\ 1988, \aj, 96, 1874.
	\bibitem[Cravens(2002)]{Cr02} Cravens, T.~E.\ 2002, Science, 296, 1042. 
	\bibitem[de Plaa et al.(2012)]{de12} de Plaa, J., Zhuravleva, I., Werner, N., et al.\ 2012, \aap, 539, A34.
	\bibitem[Dickey \& Lockman(1990)]{Di90} Dickey, J.~M. \& Lockman, F.~J.\ 1990, \araa, 28, 215. 
	\bibitem[Gabriel \& Jordan(1969)]{Ga69} Gabriel, A.~H. \& Jordan, C.\ 1969, \mnras, 145, 241.
	\bibitem[Gaensler et al.(2003)]{Ga03} Gaensler, B.~M., Hendrick, S.~P., Reynolds, S.~P., et al.\ 2003, \apjl, 594, L111. 
	\bibitem[Ghavamian et al.(2003)]{Gh03} Ghavamian, P., Rakowski, C.~E., Hughes, J.~P., et al.\ 2003, \apj, 590, 833.
	\bibitem[Haberl et al.(2012)]{Ha12} Haberl, F., Filipovi{\'c}, M.~D., Bozzetto, L.~M., et al.\ 2012, \aap, 543, A154. 
	\bibitem[Hester \& Cox(1986)]{He86} Hester, J.~J. \& Cox, D.~P.\ 1986, \apj, 300, 675.
	\bibitem[Hitomi Collaboration et al.(2018)]{Hi18} Hitomi Collaboration, Aharonian, F., Akamatsu, H., et al.\ 2018, \pasj, 70, 10.
	\bibitem[Hughes et al.(1998)]{Hu98} Hughes, J.~P., Hayashi, I., \& Koyama, K.\ 1998, \apj, 505, 732.
	\bibitem[Hughes et al.(2000)]{Hu00} Hughes, J.~P., Rakowski, C.~E., \& Decourchelle, A.\ 2000, \apjl, 543, L61.
	\bibitem[Hughes et al.(2006)]{Hu06} Hughes, J.~P., Rafelski, M., Warren, J.~S., et al.\ 2006, \apjl, 645, L117.
	\bibitem[Kaastra \& Mewe(1995)]{Ka95} Kaastra, J.~S. \& Mewe, R.\ 1995, \aap, 302, L13
	\bibitem[Kaastra et al.(1996)]{Ka96} Kaastra, J.~S., Mewe, R., \& Nieuwenhuijzen, H.\ 1996, UV and X-ray Spectroscopy of Astrophysical and Laboratory Plasmas, 411
	\bibitem[Katsuda et al.(2011)]{Ka11} Katsuda, S., Tsunemi, H., Mori, K., et al.\ 2011, \apj, 730, 24. 
	\bibitem[Katsuda et al.(2012)]{Ka12} Katsuda, S., Tsunemi, H., Mori, K., et al.\ 2012, \apj, 756, 49. 
	\bibitem[Kim et al.(2003)]{Ki03} Kim, S., Staveley-Smith, L., Dopita, M.~A., et al.\ 2003, \apjs, 148, 473.
	\bibitem[Laki{\'c}evi{\'c} et al.(2015)]{La15} Laki{\'c}evi{\'c}, M., van Loon, J.~T., Meixner, M., et al.\ 2015, \apj, 799, 50.
	\bibitem[Lallement(2004)]{La04} Lallement, R.\ 2004, \aap, 422, 391.
	\bibitem[Law et al.(2020)]{La20} Law, C.~J., Milisavljevic, D., Patnaude, D.~J., et al.\ 2020, \apj, 894, 73.
	\bibitem[Levenson et al.(1998)]{Le98} Levenson, N.~A., Graham, J.~R., Keller, L.~D., et al.\ 1998, \apjs, 118, 541.
	\bibitem[Long et al.(2014)]{Lo14} Long, K.~S., Bamba, A., Aharonian, F., et al.\ 2014, arXiv:1412.1166
	\bibitem[Lopez et al.(2009)]{Lo09} Lopez, L.~A., Ramirez-Ruiz, E., Badenes, C., et al.\ 2009, \apjl, 706, L106. 
	\bibitem[Lopez et al.(2011)]{Lo11} Lopez, L.~A., Ramirez-Ruiz, E., Huppenkothen, D., et al.\ 2011, \apj, 732, 114.
	\bibitem[McEntaffer et al.(2012)]{Mc12} McEntaffer, R.~L., Brantseg, T., \& Presley, M.\ 2012, \apj, 756, 17. 
	\bibitem[Miyata et al.(2008)]{Mi08} Miyata, E., Masai, K., \& Hughes, J.~P.\ 2008, \pasj, 60, 521. 
	\bibitem[Park et al.(2003)]{Pa03} Park, S., Burrows, D.~N., Garmire, G.~P., et al.\ 2003, \apj, 586, 210.
%	\bibitem[Park et al.(2012)]{Pa12} Park, S., Hughes, J.~P., Slane, P.~O., et al.\ 2012, \apj, 748, 117.
	\bibitem[Petre et al.(1982)]{Pe82} Petre, R., Kriss, G.~A., Winkler, P.~F., et al.\ 1982, \apj, 258, 22.
	\bibitem[Pietrzy{\'n}ski et al.(2013)]{Pi13} Pietrzy{\'n}ski, G., Graczyk, D., Gieren, W., et al.\ 2013, \nat, 495, 76. 
	\bibitem[Rasmussen et al.(2001)]{Ra01} Rasmussen, A.~P., Behar, E., Kahn, S.~M., et al.\ 2001, \aap, 365, L231.
	\bibitem[Russell \& Dopita(1992)]{Ru92} Russell, S.~C. \& Dopita, M.~A.\ 1992, \apj, 384, 508. 
	\bibitem[Sano et al.(2015)]{Sa15} Sano, H., Fukui, Y., Yoshiike, S., et al.\ 2015, Revolution in Astronomy with ALMA: The Third Year, 499, 257
	\bibitem[Sano et al.(2017)]{Sa17} Sano, H., Fujii, K., Yamane, Y., et al.\ 2017, 6th International Symposium on High Energy Gamma-Ray Astronomy, 1792, 040038.
	\bibitem[Sano et al.(2020)]{Sa20} Sano, H., Plucinsky, P.~P., Bamba, A., et al.\ 2020, \apj, 902, 53.
	\bibitem[Schenck et al.(2016)]{Sc16} Schenck, A., Park, S., \& Post, S.\ 2016, \aj, 151, 161.
	\bibitem[Someya et al.(2010)]{So10} Someya, K., Bamba, A., \& Ishida, M.\ 2010, \pasj, 62, 1301.
	\bibitem[Suzuki et al.(2020)]{Su20} Suzuki, H., Yamaguchi, H., Ishida, M., et al.\ 2020, \apj, 900, 39. 
	\bibitem[Tuohy \& Dopita(1983)]{Tu83} Tuohy, I.~R. \& Dopita, M.~A.\ 1983, \apjl, 268, L11.
	\bibitem[Uchida et al.(2015)]{Uc15} Uchida, H., Koyama, K., \& Yamaguchi, H. \ 2015, \apj, 808, 77. 
	\bibitem[Uchida et al.(2019)]{Uc19} Uchida, H., Katsuda, S., Tsunemi, H., et al.\ 2019, \apj, 871, 234. 
	\bibitem[van der Heyden et al.(2003)]{He03} van der Heyden, K.~J., Bleeker, J.~A.~M., Kaastra, J.~S., et al.\ 2003, \aap, 406, 141. 
	\bibitem[Vogt \& Dopita(2011)]{Vo11} Vogt, F. \& Dopita, M.~A.\ 2011, \apss, 331, 521.
	\bibitem[Wachter et al.(1979)]{Wa79} Wachter, K., Leach, R., \& Kellogg, E.\ 1979, \apj, 230, 274.
	\bibitem[Williams et al.(2006)]{Wi06} Williams, B.~J., Borkowski, K.~J., Reynolds, S.~P., et al.\ 2006, \apjl, 652, L33. 
	\bibitem[Winkler et al.(1988)]{Wi88} Winkler, P.~F., Tuttle, J.~H., Kirshner, R.~P., et al.\ 1988, IAU Colloq. 101: Supernova Remnants and the Interstellar Medium, 65
	\bibitem[Xu et al.(2002)]{Xu02} Xu, H., Kahn, S.~M., Peterson, J.~R., et al.\ 2002, \apj, 579, 600.
	\bibitem[Yamane et al.(2018)]{Ya18} Yamane, Y., Sano, H., van Loon, J.~T., et al.\ 2018, \apj, 863, 55.
\end{thebibliography}
\end{document}